\documentclass[cernpreprint, atlasdraft=false, paper=letter, texlive=2023, UKenglish, texmf, orcidlogo]{atlasdoc}
 
\usepackage{atlaspackage}
\usepackage{atlasbiblatex}
 
\usepackage{atlasphysics}
\usepackage{multirow}

\graphicspath{{logos/}{figures/}}

\AtlasTitle{Search for flavour-changing neutral-current couplings between the top quark and the photon with the ATLAS detector at $\sqrt{s} = 13$~TeV}

\AtlasAbstract{
This letter documents a search for flavour-changing neutral currents (FCNCs), which are strongly suppressed in the Standard Model, in events with a photon and a top quark with the ATLAS detector. The analysis uses data collected in $pp$ collisions at $\sqrt{s} = 13$~TeV during Run~2 of the LHC, corresponding to an integrated luminosity of 139~fb$^{-1}$. Both FCNC top-quark production and decay are considered. The final state consists of a charged lepton, missing transverse momentum, a $b$-tagged jet, one high-momentum photon and possibly additional jets. A multiclass deep neural network is used to classify events either as signal in one of the two categories, FCNC production or decay, or as background. No significant excess of events over the background prediction is observed and 95\% CL upper limits are placed on the strength of left- and right-handed FCNC interactions.
The 95\% CL bounds on the branching fractions for the FCNC top-quark decays, estimated (expected) from both top-quark production and decay, are $\mathcal{B}(t\rightarrow u\gamma) < 0.85\, (0.88^{+0.37}_{-0.25}) \times 10^{-5}$ and $\mathcal{B}(t\to c\gamma) < 4.2\, (3.40^{+1.35}_{-0.95})\times 10^{-5}$ for a left-handed $tq\gamma$ coupling, and $\mathcal{B}(t\to u\gamma) < 1.2\, (1.20^{+0.50}_{-0.33}) \times 10^{-5}$ and $\mathcal{B}(t\to c\gamma) < 4.5\, (3.70^{+1.47}_{-1.03}) \times 10^{-5}$ for a right-handed coupling.}
 
\AtlasRefCode{TOPQ-2019-19}
 
\PreprintIdNumber{CERN-EP-2022-042}

\AtlasJournalRef{Phys.\ Lett.\ B. 842 (2023) 137379}
 
\AtlasDOI{10.1016/j.physletb.2022.137379}

\hypersetup{pdftitle={ATLAS document},pdfauthor={The ATLAS Collaboration}}
 
\newcommand{\egfake}{\ensuremath{f_{e\rightarrow\gamma}}}
 
\newcommand{\AtlasCoordFootnote}{
ATLAS uses a right-handed coordinate system with its origin at the nominal interaction point (IP)
in the centre of the detector and the \(z\)-axis along the beam pipe.
The \(x\)-axis points from the IP to the centre of the LHC ring,
and the \(y\)-axis points upwards.
Cylindrical coordinates \((r,\phi)\) are used in the transverse plane,
\(\phi\) being the azimuthal angle around the \(z\)-axis.
The pseudorapidity is defined in terms of the polar angle \(\theta\) as \(\eta = -\ln \tan(\theta/2)\).
Angular distance is measured in units of \(\Delta R \equiv \sqrt{(\Delta\eta)^{2} + (\Delta\phi)^{2}}\).}
 
\begin{document}
 
\maketitle

\section{Introduction}

Flavour-changing neutral currents (FCNCs) are forbidden at tree level in the Standard Model (SM) and strongly suppressed at higher orders via the GIM mechanism~\cite{Glashow:1970gm}, but several extensions to the SM include additional sources of FCNCs.
In particular, some of these models predict the branching fractions (BRs) of top-quark decays via FCNCs to be orders of magnitude larger~\cite{AguilarSaavedra:2004wm} than those predicted by the SM, which are of the order of $10^{-14}$~\cite{AguilarSaavedra:2004wm}.
Examples are R-parity-violating supersymmetric models~\cite{Yang:1997dk,Delepine:2004hr,Cao:2007dk,Yang:2018utw} and models with two Higgs doublets~\cite{Atwood:1996vj,Bejar:2000ub}, which allow FCNC processes involving top quarks to have measurable rates.
 
\begin{figure}[ht!]
\begin{center}
\begin{tabular}{cc}
\includegraphics[width=0.47\columnwidth]{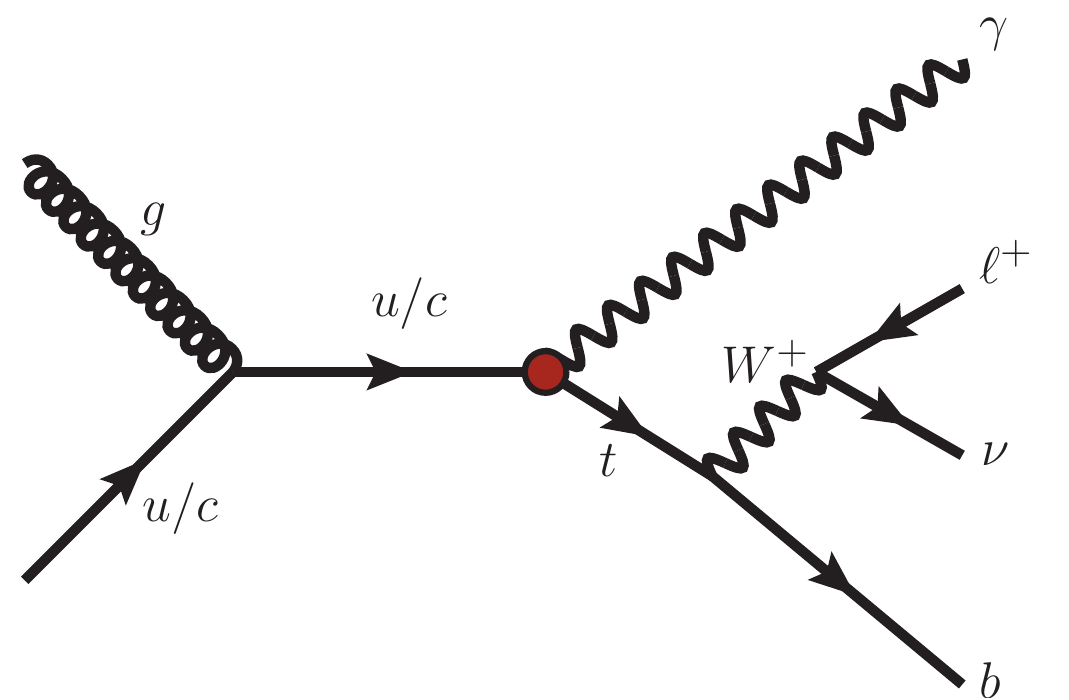} &
\includegraphics[width=0.47\columnwidth]{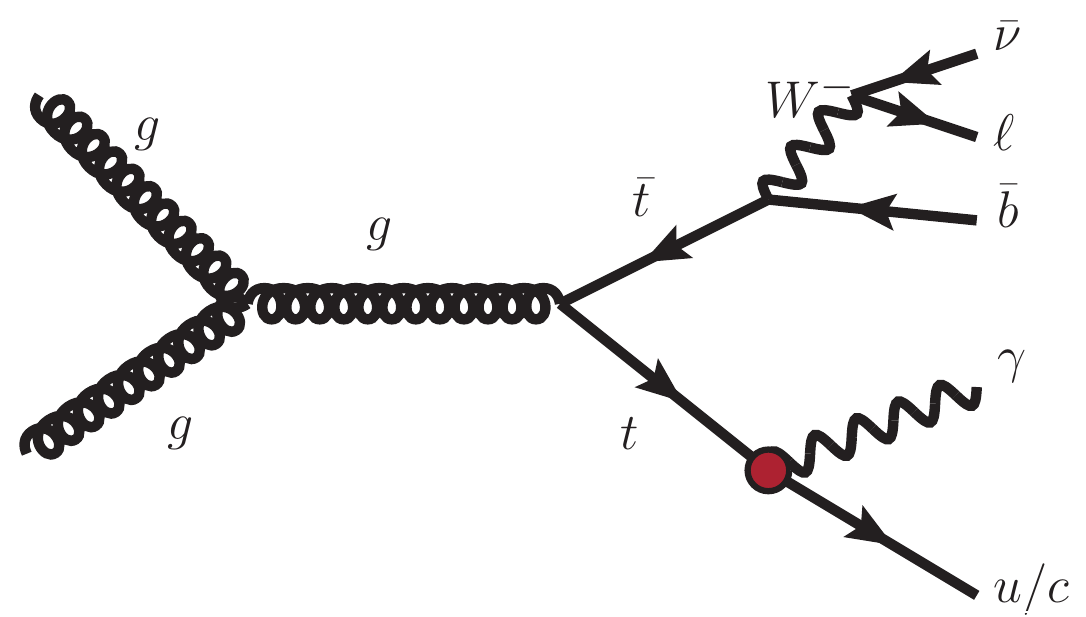}
\end{tabular}
\caption{Tree-level Feynman diagrams for top-quark production (left) and decay (right) via FCNCs. The $tq\gamma$ vertex, which is not present in the SM, is highlighted.
\label{fig:feynman}}
\end{center}
\end{figure}

This letter presents a search for FCNCs in processes with a top quark ($t$) and a photon ($\gamma$) based on the full dataset of $\sqrt{s} = 13~\tev$ proton--proton collisions collected by the ATLAS experiment~\cite{PERF-2007-01} during Run~2 of the LHC.
The analysis is optimised to search for the production of a single top quark in association with a photon as well as to search for the decay of a top quark into an up ($u$) or charm ($c$) quark in association with a photon in the case of pair-produced top quarks ($t\bar{t}$).
Tree-level Feynman diagrams for these processes are shown in Figure~\ref{fig:feynman}, where in both cases, exactly one top quark decays via the SM-favoured $tWb$ coupling.
 
FCNC contributions to the production ($q \rightarrow t \gamma$, $q= u,c$, Figure~\ref{fig:feynman} left), and decay ($t\rightarrow q \gamma$, Figure~\ref{fig:feynman} right), modes can be parameterised in terms of effective coupling parameters~\cite{AguilarSaavedra:2008zc,Durieux:2014xla}.
Following the notation of Refs.~\cite{Grzadkowski:2010es,Aguilar-Saavedra:2018ksv}, the relevant dimension-six operators are $O_{\mathrm{uB}}^{(ij)}$ and $O_{\mathrm{uW}}^{(ij)}$, where $i\neq j$ are indices for the quark generation.
In general, left-handed (LH) and right-handed (RH) couplings could exist, resulting in different helicities for the top quark in the production mode, which leads to different kinematic properties for the final-state particles from the weak decay of the top quark.
 
The CMS Collaboration has searched for the production mode using data taken at $\sqrt{s} = 8~\tev$~\cite{Khachatryan:2015att}.
The ATLAS Collaboration also performed an analysis that was optimised for the production mode using 81~fb$^{-1}$ of data at $\sqrt{s} = 13~\tev$~\cite{TOPQ-2018-22}, resulting in the strongest upper limits to date.
The limits on the LH (RH) effective coupling parameters were translated into BR upper limits of $2.8 \times 10^{-5}$ ($6.1 \times 10^{-5}$) for $t\rightarrow u \gamma$ and $22 \times 10^{-5}$ ($18 \times 10^{-5}$) for $t\rightarrow c \gamma$.
The search presented in this letter supersedes the one in Ref.~\cite{TOPQ-2018-22}, uses the full 139~fb$^{-1}$ Run~2 dataset at $\sqrt{s}=13~\tev$ and is optimised for both the decay and production modes by training a neural-network (NN) classifier to separate the signal in decay and production modes from the SM background.


\section{The ATLAS detector}

ATLAS~\cite{PERF-2007-01,Abbott:2018ikt,CERN-LHCC-2012-009} is a multipurpose particle detector designed with a forward--backward symmetric cylindrical geometry and nearly full \(4\pi\) coverage in solid angle.\footnote{\AtlasCoordFootnote}
It consists of an inner tracking detector (ID) surrounded by a thin superconducting solenoid providing a \SI{2}{\tesla} axial magnetic field, electromagnetic and hadronic calorimeters, and a muon spectrometer (MS).
The ID covers the pseudorapidity range \(|\eta| < 2.5\) and is composed of silicon pixel, silicon microstrip, and transition radiation tracking (TRT) detectors.
Lead/liquid-argon (LAr) sampling calorimeters provide electromagnetic (EM) energy measurements with high granularity.
Hadronic calorimetry is provided by the steel/scintillator-tile calorimeter covering the central pseudorapidity range (\(|\eta| < 1.7\)).
The endcap and forward regions are instrumented with LAr calorimeters for both the EM and hadronic energy measurements up to \(|\eta| = 4.9\).
The MS surrounds the calorimeters and is based on three large air-core toroidal superconducting magnets with eight coils each.
The field integral of the toroids ranges between \num{2.0} and \SI{6.0}{\tesla\metre} across most of the detector.
The MS includes a system of precision tracking chambers and fast detectors for triggering.
A two-level trigger system is used to select events.
The first-level trigger is implemented in hardware and uses a subset of the detector information
to keep the accepted event rate below \SI{100}{\kHz}~\cite{TRIG-2016-01}.
This is followed by a software-based trigger that reduces the accepted event rate to \SI{1}{\kHz} on average.
An extensive software suite~\cite{ATL-SOFT-PUB-2021-001} is used in the reconstruction and analysis of real
and simulated data, in detector operations, and in the trigger and data acquisition systems of the experiment.


\section{Analysis strategy}

A signal region (SR) is defined by selecting events that contain one high-momentum photon and the decay products of a semileptonically decaying top quark, i.e.\ an electron or muon, a $b$-tagged jet and missing transverse momentum.
Additional jets may be present in the final state, as one such jet is expected at leading order for signal in the decay mode and additional jets may result from initial- or final-state radiation in both signal modes.
The main backgrounds stem from events with prompt photons (mostly $\ttbar\gamma$ events in the lepton+jets channel and $W\gamma$+jets events), from events with an electron that is misidentified as a photon (referred to as $e\to\gamma$ fakes, mostly in dileptonic \ttbar\ events), and from events with hadrons that are misidentified as photons (referred to as $h\to\gamma$ fakes, mostly in semileptonic \ttbar events).
Backgrounds with prompt photons are modelled by Monte Carlo (MC) simulations, and control regions (CRs) are defined for the $\ttbar\gamma$ and the $W\gamma$+jets processes.
The $\ttbar\gamma$ CR is based on the presence of additional jets, especially an additional $b$-tagged jet.
The $W\gamma$+jets CR is constructed by requiring that the $b$-tagged jet fulfils only a looser $b$-tagging requirement and not the tight $b$-tagging requirement that is used in the SR.
The contributions from $e\to\gamma$ and $h\to\gamma$ fakes are modelled by MC simulations but are corrected with data-driven scale factors (SFs).
In the SR, signal and background events are distinguished using a NN with three output nodes, one for each signal mode and one for the SM background.
The three nodes are combined into a one-dimensional discriminant to separate the total signal from the background.
The signal contribution is then estimated with a binned profile likelihood fit to this discriminant, with systematic uncertainties modelled as nuisance parameters.
Separate NNs are trained for the $tu\gamma$ and $tc\gamma$ couplings because the signal processes differ in their kinematic properties, due to differences between the up- and charm-quark parton distribution functions (PDFs), and also in their $b$-tagging probabilities.
The different $b$-tagging properties stem from the jets initiated by the $c$-quark from the FCNC top-quark decay in the case of the $tc\gamma$ coupling.


\section{Data and simulation}
\label{sec:samples}

The proton--proton ($pp$) collision data analysed for this search were recorded with the ATLAS detector from 2015 to 2018 at a centre-of-mass energy of $\sqrt{s} = 13$~\TeV.
Events were selected using single-lepton triggers~\cite{TRIG-2016-01,TRIG-2018-05,TRIG-2018-01} and are required to have at least one reconstructed primary vertex that has at least three associated tracks with transverse momenta greater than 500~\MeV.
After the application of data-quality requirements~\cite{Aad:2019hmz}, the data sample corresponds to an integrated luminosity of 139~fb$^{-1}$, as determined by using the LUCID-2 detector~\cite{LUCID2} for the primary luminosity measurements.
 
Monte Carlo simulated events samples are used in the analysis to optimise the event selection, to train the NN and to predict contributions from various SM processes.
The effect of multiple interactions in the same and neighbouring bunch crossings (pile-up) was modelled by overlaying the simulated hard-scattering event with inelastic $pp$ events generated by \PYTHIA[8.186]~\cite{Sjostrand:2007gs} using the \NNPDF[2.3lo] set of PDFs~\cite{Ball:2012cx} and parameter values set according to the A3 tune~\cite{ATL-PHYS-PUB-2016-017}.
After the event generation, the ATLAS detector response was simulated~\cite{SOFT-2010-01,} by using the \GEANT toolkit~\cite{Agostinelli:2002hh} with either the full simulation of the ATLAS detector or the fast-simulation package~\cite{ATL-PHYS-PUB-2010-013}.
In all processes, the top-quark mass was set to $172.5~\gev$.
For all samples of simulated events, except those generated using \SHERPA, the decays of bottom and charm hadrons were performed by \EVTGEN~\cite{Lange:2001uf}.
 
The signal in the production mode $(pp \to t\gamma \to b\bar{\ell}\nu\gamma)$ and in the decay mode $(pp \to \ttbar \to b \bar{\ell}\nu\bar{q}\gamma)$ was simulated with the \MGNLO[2.4.3] generator~\cite{Mrenna:2016sih} with the UFO model TopFCNC~\cite{PhysRevD.91.034024,Durieux:2014xla} at next-to-leading order (NLO) in QCD, using the \NNPDF[3.0nlo] PDF set~\cite{Ball:2014uwa}.
The scale of new physics was set to $\Lambda = 1~\tev$.
The charged leptons, $\ell$, include electrons, muons and tau leptons.
Interference effects between the production and decay FCNC processes can be neglected~\cite{Barros:2019wxe}.
For SM decays of the top quark, the spin correlation was preserved using \MADSPIN~\cite{Artoisenet:2012st}.
The parton showering and hadronisation were simulated using \PYTHIA[8.212]~\cite{Sjostrand:2014zea} with the A14 tune~\cite{ATL-PHYS-PUB-2014-021} and the \NNPDF[2.3lo] PDF set.
In the production mode, four samples were generated with different couplings set to non-zero values: $tu\gamma$ LH, $tu\gamma$ RH, $tc\gamma$ LH, and $tc\gamma$ RH.
In the decay mode, only samples with $tu\gamma$ LH coupling and $tc\gamma$ LH coupling have been simulated since the kinematic properties of the LH and RH couplings were found to be very similar.
Correspondingly, only the LH coupling for the $tc\gamma$ coupling in the decay is considered.
The production and decay processes contribute similarly for the $tu\gamma$ coupling, while for the $tc\gamma$ coupling the decay is the dominant process.
For given values of the Wilson coefficients, the cross-section for the production process is calculated with \MGNLO at NLO using the TopFCNC model.
Then, for given values of the Wilson coefficients, the $tq\gamma$ BR is calculated by the LO relation between the BR and the Wilson coefficients as given in Ref.~\cite{Durieux:2014xla}.
The following effective cross-sections, calculated as total cross-section times branching fractions assuming $\mathcal{B}(t\to q\gamma)$ of $10^{-3}$, are used for the production processes: 417~fb for the $tu\gamma$ LH coupling, 416~fb for the $tu\gamma$ RH coupling, 59.5~fb for both the $tc\gamma$ LH coupling and the $tc\gamma$ RH coupling.
For the decay processes, the effective cross-section is 542~fb for both $tu\gamma$ and $tc\gamma$ couplings.
 
Two dedicated MC samples are used to model $\ttbar\gamma$ events.
In the first sample, photon radiation in \ttbar\ production was generated at NLO precision in QCD.
In the second sample, photons in the decay were generated at LO precision in QCD while removing the overlap with the former sample.
Events in both samples were modelled using the \MGNLO[2.7.3]~\cite{Alwall:2014hca} generator with the \NNPDF[2.3lo] PDF set.
The events were interfaced with \PYTHIA[8.240]~\cite{Sjostrand:2014zea}, which used the A14 tune and the \NNPDF[2.3lo] PDF set.
To combine the two samples, a $K$-factor is applied to scale the predicted cross-section of the decay sample, while no $K$-factor is applied to the production sample.
The nominal value of the $K$-factor for the decay sample is chosen so that the inclusive cross-section agrees with the theoretical cross-section calculation at  NLO in QCD in Ref.~\cite{Melnikov:2011ta}, resulting in a $K$-factor of 1.67 for the decay sample.
 
The production of \ttbar\ events was modelled using the \POWHEGBOX[v2]~\cite{Frixione:2007nw,Nason:2004rx,Frixione:2007vw,Alioli:2010xd} generator at NLO with the \NNPDF[3.0nlo] PDF set
and the \hdamp\ parameter\footnote{The \hdamp\ parameter is a resummation damping factor and one of the parameters that controls the matching of \POWHEG matrix elements to the parton shower and thus effectively regulates the high-\pt\ radiation against which the \ttbar\ system recoils.} set to 1.5~times the top-quark mass~\cite{ATL-PHYS-PUB-2016-020}.
The events were interfaced to \PYTHIA[8.230]~\cite{Sjostrand:2014zea} to model the parton shower, hadronisation, and underlying event, with parameters set according to the A14 tune and using the \NNPDF[2.3lo] set of PDFs.
The \ttbar\ sample is normalised to the cross-section prediction at next-to-next-to-leading order (NNLO) in QCD including the resummation of next-to-next-to-leading logarithmic (NNLL) soft-gluon terms calculated using \TOPpp[2.0]~\cite{Beneke:2011mq,Cacciari:2011hy,Baernreuther:2012ws,Czakon:2012zr,Czakon:2012pz,Czakon:2013goa,Czakon:2011xx}.
 
Simulated events for the SM $tq\gamma$ process, with photon radiation in the production of the top quark, were generated at NLO in the four-flavour scheme using the \MGNLO[2.6.2] event generator interfaced with \PYTHIA[8.240] for parton showering.
Contributions with photon radiation from top-quark decay products were modelled by using the SM single-top-quark $t$-channel sample described below.
The top quark was decayed using~\MADSPIN.
 
The single-top-quark samples are split into three processes: $s$-channel, $t$-channel and $tW$-channel.
These samples were modelled using the \POWHEGBOX[v2]~\cite{Frederix:2012dh} generator at NLO in QCD using the four-flavour (five-flavour) scheme for the $t$-channel ($s$-channel and $tW$-channel) and the corresponding \NNPDF[3.0nlo] set of PDFs.
In the case of the $tW$-channel, the diagram removal scheme~\cite{Frixione:2008yi} was used.
To avoid an overlap with the SM $tq\gamma$ sample, $t$-channel events with a prompt photon were only selected when the photon is radiated from the top-quark decay products.
The events were interfaced with \PYTHIA[8.230], which used the A14 tune and the \NNPDF[2.3lo] set of PDFs.
 
The production of $V\gamma$+jets ($V=W,Z$) final states was simulated with the \SHERPA[2.2.8]~\cite{Bothmann:2019yzt} generator.
Matrix elements (MEs) at NLO QCD accuracy for up to one additional parton and at LO accuracy for up to three additional parton emissions were matched and merged with the \SHERPA parton shower based on Catani--Seymour dipole factorisation~\cite{Gleisberg:2008fv,Schumann:2007mg} using the MEPS@NLO prescription~\cite{Hoeche:2011fd,Hoeche:2012yf,Catani:2001cc,Hoeche:2009rj}.
The virtual QCD corrections for MEs at NLO accuracy were provided by the \OPENLOOPS library~\cite{Buccioni:2019sur,Cascioli:2011va,Buccioni:2017yxi,Denner:2016kdg}.
Samples were generated using the \NNPDF[3.0nlo] set of PDFs, along with the dedicated set of tuned parton-shower parameters developed by the \SHERPA authors.
 
The production of $V\!+$jets events was simulated with the \SHERPA[2.2.1]~\cite{Bothmann:2019yzt} generator using NLO MEs for up to two partons, and LO MEs for up to four partons, calculated with the Comix~\cite{Gleisberg:2008fv} and \OPENLOOPS libraries.
The MEPS@NLO prescription was used to match the MEs with the \SHERPA parton shower, which used the set of tuned parameters developed by the \SHERPA authors.
The \NNPDF[3.0nlo] set of PDFs was used and the samples are normalised to a NNLO prediction~\cite{Anastasiou:2003ds}.
 
Samples of diboson final states ($VV$) were simulated with the \SHERPA[2.2.1] or 2.2.2~\cite{Bothmann:2019yzt} generator depending on the process, including off-shell effects and Higgs boson contributions where appropriate.
Fully leptonic final states and semileptonic final states, where one boson decays leptonically and the other hadronically, were generated using MEs at NLO accuracy in QCD for up to one additional parton and at LO accuracy for up to three additional parton emissions.
The \NNPDF[3.0nlo] set of PDFs was used, along with the dedicated set of tuned parton-shower parameters developed by the \SHERPA authors.
 
An overlap removal scheme was applied to remove double-counting of events stemming from photon radiation in samples in which a photon was not explicitly required in the final state, similarly to Ref.~\cite{TOPQ-2017-14}.
This procedure was applied to the $\ttbar$, $W$\!+jets, $Z$+jets and single-top $t$-channel samples, to avoid overlaps with the $\ttbar\gamma$, $W\gamma$+jets, $Z\gamma$+jets and SM $tq\gamma$ samples, respectively.


\section{Object and event selection}
\label{sec:selection}

Electron candidates are reconstructed from clusters of energy deposits in the electromagnetic calorimeter matched to charged-particle tracks in the ID.
The candidates must satisfy the \emph{TightLH} likelihood-based identification criteria~\cite{ATL-PHYS-PUB-2017-022,EGAM-2018-01} with $\pt > 27~\gev$ and $|\eta| < 2.47$, with the region $1.37 < |\eta| < 1.52$ excluded.
Additionally, electron candidates must meet two impact-parameter selection criteria: $|z_0 \sin{\theta}| < 0.5$~mm, where $z_0$ is the $z$ coordinate of the transverse impact point, and $|d_0/\sigma(d_0)| < 5$ for the transverse impact-parameter significance.
Muon candidates are reconstructed from tracks in the MS matched to tracks in the ID.
The candidates are required to meet \emph{Medium} identification criteria~\cite{MUON-2018-03} with $\pt > 27~\gev$ and $|\eta| < 2.5$.
Additionally, muon candidates must satisfy $|z_0 \sin{\theta}| < 0.5$~mm and $|d_0/\sigma(d_0)| < 3$.
Isolated electrons and muons are selected by requiring both the amount of energy deposited nearby in the calorimeters and the scalar sum of the transverse momenta of nearby tracks in the ID to be small.
 
Photon candidates are reconstructed from clusters of energy deposits in the electromagnetic calorimeter that have either no matched ID track or one or two matched ID tracks which are compatible with electron or positron tracks from a photon conversion~\cite{EGAM-2018-01}.
Based on the number of ID tracks matched to the electromagnetic cluster, photons are separated into different categories.
If one or two tracks are matched, the photon is labelled as converted; if zero tracks are matched, the photon is labelled as unconverted.
Photon candidates are required to have $\pt > 20~\gev$ and be within $|\eta| < 2.37$, with the region at $1.37 < |\eta| < 1.52$ excluded.
The candidates are required to meet the \emph{Tight} identification criteria~\cite{ATL-PHYS-PUB-2017-022,EGAM-2018-01}.
Photons must be isolated from nearby energy deposits in the calorimeter and from nearby tracks in the ID.
The sum of the energy deposited ($\pt$ of the tracks) within $\Delta R = 0.4$ ($\Delta R = 0.2$) of the photon direction is required to be smaller than $0.022\times\pTX[\gamma] + 2.45~\GeV$ ($0.05\times\pTX[\gamma]$), excluding the photon energy deposition (tracks associated with the photon)~\cite{EGAM-2018-01}.
 
Jet candidates are reconstructed from particle-flow objects~\cite{PERF-2015-09}, using the \antikt~\cite{Cacciari:2008gp} jet algorithm with radius parameter $R=0.4$ implemented in FastJet~\cite{Fastjet}.
Jets are calibrated by applying a jet energy scale derived from 13~\TeV\ data and simulation~\cite{JETM-2018-05}.
After the calibration, jet candidates are required to have $\pt > 25~\gev$ and $|\eta| < 2.5$.
To suppress jets originating from pile-up collisions, a cut on the jet-vertex tagger (JVT)~\cite{PERF-2014-03} discriminant as defined in the \emph{JVTTight} working point (WP) is applied for jets with \pt below 120~\gev.
Jets containing $b$-hadrons are identified ($b$-tagged) using a deep neural network.
This neural network, DL1r~\cite{FTAG-2018-01,ATL-PHYS-PUB-2017-013}, combines inputs from the impact parameters of tracks and the displaced vertices reconstructed in the ID.
The inputs of the DL1r neural network also include discriminating variables constructed by a recurrent neural network, which exploits the spatial and kinematic correlations between tracks originating from the same $b$-hadron.
The outputs of the neural network represent the probabilities of the jet to originate from a light-flavour quark or gluon, a $c$-quark and a $b$-quark, which are then combined into a single discriminant.
 
The missing transverse momentum vector $\vec{p}_{\textrm{T}}^{\text{\,miss}}$, with magnitude \met, is defined as the negative sum of the transverse momenta of the reconstructed and calibrated physical objects and a soft term built from all tracks that are associated with the primary vertex but not with these objects~\cite{PERF-2016-07}.
 
To avoid double-counting of detector signatures, objects are removed in the following order:\footnote{For the overlap removal, $\Delta R$ is defined as $\Delta R \equiv \sqrt{\left(\Delta y\right)^2 + \left(\Delta \phi\right)^2}$, where $y$ is the rapidity of the object.}
electrons sharing a track with a muon;
jets within $\Delta R = 0.2$ of an electron;
electrons within $\Delta R = 0.4$ of a jet;
jets within $\Delta R = 0.4$ of a muon if they have at most two associated tracks;
muons within $\Delta R = 0.4$ of a jet;
photons within $\Delta R = 0.4$ of an electron or muon;
jets within $\Delta R = 0.4$ of a photon.
 
SFs are used to correct the efficiencies in simulation in order to match the efficiencies measured in data for the electron~\cite{EGAM-2018-01,PERF-2017-01,PERF-2017-03,PERF-2015-10} and muon~\cite{PERF-2015-10} trigger, reconstruction, identification, and isolation criteria, as well as for the photon identification~\cite{EGAM-2018-01} and isolation requirements.
SFs are also applied for the JVT requirement~\cite{ATLAS-CONF-2014-018} and for the $b$-tagging efficiencies for jets that originate from the hadronisation of $b$-quarks~\cite{FTAG-2018-01}, $c$-quarks~\cite{ATLAS-CONF-2018-001}, and $u$-, $d$-, $s$-quarks or gluons~\cite{ATLAS-CONF-2018-006}.
 
\begin{figure}[ht!]
\begin{center}
\includegraphics[width=\linewidth]{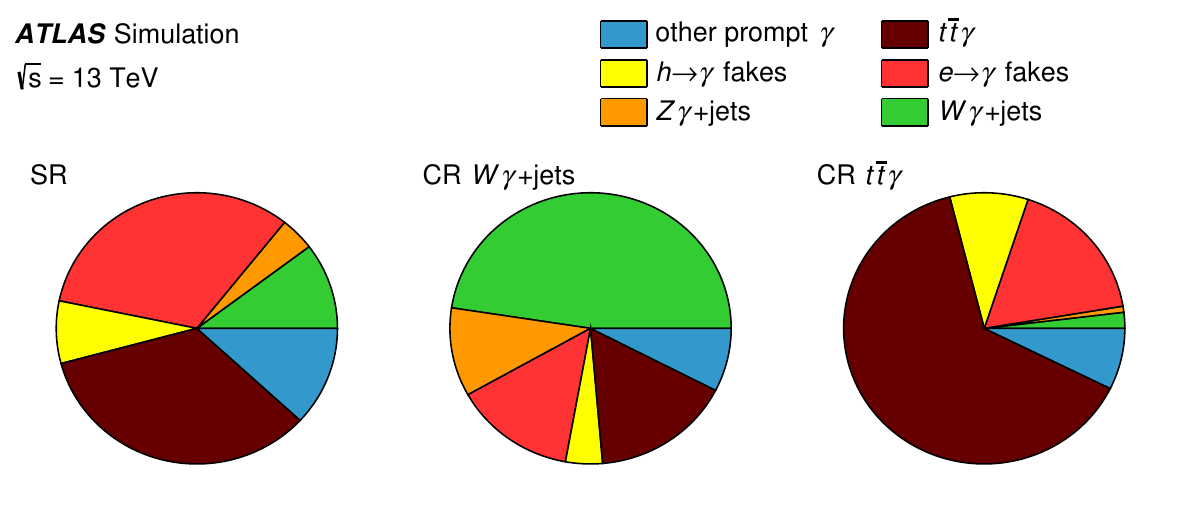}
\caption{Expected background composition of the SR (left), $W\gamma$+jets CR (middle) and $\ttbar\gamma$ CR (right).}
\label{fig:piecharts}
\end{center}
\end{figure}

The selected events have exactly one electron or muon, exactly one photon, at least one jet and $\met > 30~\GeV$.
The lepton must be matched, with $\Delta R < 0.15$, to the lepton reconstructed by the trigger.
Events meeting these criteria are further separated into three orthogonal regions: the SR, the $t\bar{t}\gamma$ CR and the $W\gamma$+jets CR.
In the SR, exactly one $b$-tagged jet identified with the 60\% efficiency WP is required while vetoing events with additional jets that pass the 77\% WP, to ensure orthogonality to the CRs.
Events in the $t\bar{t}\gamma$ CR are required to have at least four jets with at least two $b$-tagged jets identified at the 77\% WP, and at least one of these must also be $b$-tagged at the 70\% WP.
In order to select a $W\gamma$+jets event sample with jets originating from light-flavour and heavy-flavour quarks in proportions similar to those in events satisfying the SR criteria, events in the $W\gamma$+jets CR are required to have exactly one $b$-tagged jet identified at the 77\% WP, with no jet passing the 70\% WP.
Additionally, events with an electron--photon pair invariant mass of 80--100~\gev\ are excluded from the $W\gamma$+jets CR to suppress $Z$+jets events with $e\to\gamma$ fakes.
Table~\ref{tab:EventSelection/Regions} summarises the selection criteria for the individual analysis regions.
The selection efficiency of the signal events in the SR is about 8\% for the FCNC production mode and about 6.5\% for the FCNC decay mode for the $tu\gamma$ LH coupling.
The composition of the SM backgrounds after the selection is presented in Figure~\ref{fig:piecharts}.
Around 60\% (35\%) of the jets in the signal region originate from the hadronisation of $b$-quarks ($c$-quarks), while around 10\% (75\%) of the jets in the $W\gamma$+jets CR originate from the hadronisation of $b$-quarks ($c$-quarks).
The residual differences between the kinematics of jets originating from the hadronisation of $b$- and $c$-quarks were found to be negligible.
Setting the FCNC couplings to the 95\% CL limits measured in Ref.~\cite{TOPQ-2018-22}, the signal contamination in the CRs is less than 0.3\% (1\%) for the $t\bar{t}\gamma$ CR ($W\gamma$+jets CR).
 
\begin{table}[hbpt]
\centering
\caption{Summary of the analysis region definitions. While the requirements on photons, leptons and \met are shared, the regions differ in their jet and $b$-tagged jet requirements. The latter ensure orthogonality. All jets that pass the 60\% $b$-tagging WP automatically pass the looser $b$-tagging WPs. A hyphen indicates that no criterion has to be fulfilled.}
\label{tab:EventSelection/Regions}
\small
\begin{tabular}{c|c|c|c}
\toprule{Object} & {SR} & {CR $t\bar{t}\gamma$} & {CR $W\gamma$+jets} \\
\midrule
Photon ($\pt > 20~\gev$) & \multicolumn{3}{c}{$= 1$} \\
Lepton ($\pt > 27~\gev$) & \multicolumn{3}{c}{$= 1$} \\
$E_{\text{T}}^{\text{miss}}$ & \multicolumn{3}{c}{$> \SI{30}{\giga\electronvolt}$} \\
\midrule
Jets ($\pt > 25~\gev$) & $\geq 1$ & $\geq 4$ & $\geq 1$ \\
$\,b$-tagged jets (60\% WP) & $= 1$ & -- & $= 0$ \\
$\,b$-tagged jets (70\% WP) & $=1$ & $\geq 1$ & $= 0$ \\
$\,b$-tagged jets (77\% WP) & $= 1$ & $\geq 2$ & $= 1$ \\
$m(e,\gamma)$ & -- & -- & $\notin [80, 100]$~\GeV \\
\bottomrule
\end{tabular}
\end{table}
 
\FloatBarrier


\section{Data-driven estimate of misidentified photons}
\label{sec:photonfakes}

\subsection{Electrons misidentified as photons}
 
Electrons can be misidentified as photons, for example, if the electron track is not reconstructed or if the track is not matched to the energy clusters in the electromagnetic calorimeter.
These misidentified photons are referred to in the following as $e\to\gamma$ fakes.
The probability for an electron to be misidentified as a photon, $\egfake$, is measured from data and simulation following the methodology used previously~\cite{PERF-2017-02,TOPQ-2018-22}.
The ratio of the probabilities measured in data and simulation is then applied to correct the simulation.
 
Two regions are defined in order to measure $\egfake$.
The $Z\to e\gamma$ region is defined by requiring exactly one electron and one photon with an electron--photon invariant mass in the range 70--110~\gev.
The $Z \to ee$ region is defined by requiring exactly two electrons with opposite electric charge, no photons, and a dielectron invariant mass in the range 70--110~\gev.
Both regions are required to have $\met < 30~\gev$, and a veto on the presence of $b$-tagged jets is applied.
In both regions, electrons and muons need to fulfil the same criteria as described in Chapter~\ref{sec:selection}.
Templates from MC simulation for the invariant mass of the dielectron (electron--photon) pair are estimated from a binned-likelihood fit in the $Z\to ee$ ($Z \to e\gamma$) region.
The templates also include the SM backgrounds originating from $W\gamma$+jets and $Z\gamma$+jets events.
For the $Z\to e\gamma$ region, third-order Bernstein polynomials are used to create templates for the remaining SM backgrounds.
The ratio of the integrals of the aforementioned fitted signal templates is calculated in order to estimate $2 \egfake$, where the factor of two accounts for the two electrons in $Z\rightarrow ee$ events that may be misidentified as a photon.
The $\egfake$ probabilities are measured independently in six bins in photon $|\eta|$ and separately for the different reconstruction types for converted photons, defined by the number of hits in the silicon detectors and in the TRT.
The measured $\egfake$ probabilities range from about 3\% for the central photons up to 12\% for the forward photons, with an inclusive probability of about 5\%.
 
The following systematic uncertainties in the estimated $\egfake$ probabilities are assessed: removing the third-order Bernstein polynomial functions used to describe the background contributions from the fits, removing the MC templates for $W\gamma$+jets and $Z\gamma$+jets from the fits, varying the fit range from 70--110~\gev\ to 80--100~\gev, increasing and decreasing the photon energy by 1\%, and considering modelling uncertainties of the $Z$+jets events, as the $Z$+jets process is the only process that contributes significantly.
SFs are defined by the ratios of the measured and simulated values of $\egfake$, and range from about 0.8 for the forward photons to 2.5 for the central photons for one of the reconstruction types for the converted photons.
The total uncertainties differ between the $\eta$ bins and the reconstruction types and vary from about 1.5\% to about 30\%, dominated by systematic uncertainties arising from the removal of the Bernstein polynomials and from the $Z$+jets modelling uncertainties in most bins.
The validity of the measured $\egfake$ SFs was cross-checked using a selection similar to the $Z\to e\gamma$ region but requiring at least one $b$-tagged jet to be present.
Good agreement between the corrected prediction and the data was observed.
The main processes contributing to the analysis regions via $e\to\gamma$ fakes are \ttbar with about 65\% and $Z$+jets with about 25\%.
 
\subsection{Hadrons misidentified as photons}
 
A jet can be misreconstructed as a photon, e.g.\ when a hadron inside the jet decays into two photons that are reconstructed as a single one.
The number of events with misidentified hadrons is estimated from data.
SFs, defined as the ratio of the numbers of misidentified hadrons in data and in simulation, are applied to the simulation to correct the normalisation of the events with misidentified hadrons.
The shapes of the distributions for these processes are estimated from the MC simulation with their uncertainties.
 
Three hadron fake regions (HFR) are defined by the same criteria as the SR but with modified photon identification and isolation requirements following the same prescription as in Ref.~\cite{TOPQ-2018-22}.
Assuming no correlation between the identification and isolation variables used to define the hadron fake regions, the contribution of this background in the SR can be estimated  using the ABCD method (see e.g.\ Ref.~\cite{PhysRevD.44.29}) as
 
\begin{equation*}
N(\text{SR}^{\text{pass\&pass, data}}_{h\to\gamma}) = \frac{N(\text{HFR}^{\text{pass\&fail, data}})\cdot N(\text{HFR}^{\text{fail\&pass, data}})}{N(\text{HFR}^{\text{fail\&fail, data}})},
\label{eq:fakes}
\end{equation*}
 
\noindent where $N(\text{HFR}^{\text{pass\&fail, data}})$ represents the number of $h\to\gamma$ events estimated from data after subtracting contributions from other sources in the HFR where events pass the identification requirements and fail the isolation requirements, as defined in Ref.~\cite{TOPQ-2018-22}, and
$N(\text{HFR}^{\text{fail\&pass, data}})$ and $N(\text{HFR}^{\text{fail\&fail, data}})$ are defined analogously.
Since the identification and isolation variables are not fully uncorrelated, the estimate is corrected for the non-zero correlation as estimated from MC simulations.
The correction factor for the correlations ranges from about 0.9 to about 1.15 depending on the photon~\pt\ and $\eta$ bin.
The data-driven SFs are then calculated as $\text{SF}(h\rightarrow\gamma) = N(\text{SR}_{h\rightarrow\gamma}^{\text{pass\&pass, data, corr}})/N(\text{SR}_{h\rightarrow\gamma}^{\text{pass\&pass, MC, corr}})$, where $N(\text{SR}_{h\rightarrow\gamma}^{\text{pass\&pass, MC, corr}})$ represents the number of events the MC simulation predicts in the SR after the correction for the non-zero correlations of the identification and isolation variables.
The SFs are estimated independently in six photon $|\eta|$ bins and two photon \pt bins, and separately for converted and unconverted photons.
 
The following sources of systematic uncertainty are considered for the estimation of the SFs: the finite number of events in the data and MC simulated samples, variations of the $e\to\gamma$ SFs used to subtract the $e\to\gamma$ background in the different regions by one standard deviation, variations of the correlation between identification and isolation variables, estimated as the maximum correlation seen in the simulation across all \pt and $\eta$ bins, and variations in the prompt-photon subtraction in the non-tight regions by considering normalisation uncertainties for the individual processes and modelling uncertainties of the \ttbar\ and $\ttbar\gamma$ processes.
 
The measured data-driven SFs range from about 0.6 for high-\pt\ central unconverted photons to up to 2.2 for high-\pt\ forward unconverted photons.
The total uncertainties in the SFs vary between about 45\% to 65\%, where in most of the bins the dominant uncertainty arises from the assumptions about the correlation of the photon identification and isolation variables.
The main processes contributing to the analysis regions via $h\to\gamma$ fakes are \ttbar with about 80\% and single top with about 10\%.


\section{Neural network for discrimination between signal and background}
\label{sec:NN}

The signal is distinguished from the sum of the background processes by a fully connected feed-forward NN with backpropagation, implemented in {\textsc Keras}~\cite{chollet2015keras} with the {\textsc TensorFlow}~\cite{tensorflow2015-whitepaper} back end.
Separate NNs are trained for FCNC processes with a $tu\gamma$ or a $tc\gamma$ vertex, reflecting the differences stemming mainly from the different PDFs involved in the production mode.
Differences between the LH and RH couplings were found to mostly impact the acceptance of the event selection.
The following acceptances were found for the FCNC production modes: $8.2\%$ for the $tu\gamma$ LH coupling, $6.7\%$ for the $tu\gamma$ RH coupling, $9.3\%$ for the $tc\gamma$ LH coupling and $7.5\%$ for the $tc\gamma$ RH coupling.
For the decay modes, the acceptances are $6.6\%$ for the $tu\gamma$ coupling and $5.8\%$ for the $tc\gamma$ coupling.
However, no significant impact on the discrimination power of the network was found originating from the difference of the LH and RH couplings.
Thus, the LH and RH couplings are not separated in the network.
 
An optimised set of 37 variables is used as the input to the NN: these were selected by removing the input variables with negligible impact on the separation power of the final discriminant.
The input variables include the \pt and $\eta$ of the charged leptons, photons, $b$-tagged jets and the two leading non-$b$-tagged jets, \met and the photon conversion status.
High-level variables, such as invariant masses and angular distances between the objects, jet multiplicities and the $b$-tagging information, are also included.
All variables are transformed using {\textsc scikit-learn}'s~\cite{Pedregosa:2012toh} StandardScaler.
 
\begin{figure}[ht!]
\begin{center}
\includegraphics[width=\linewidth]{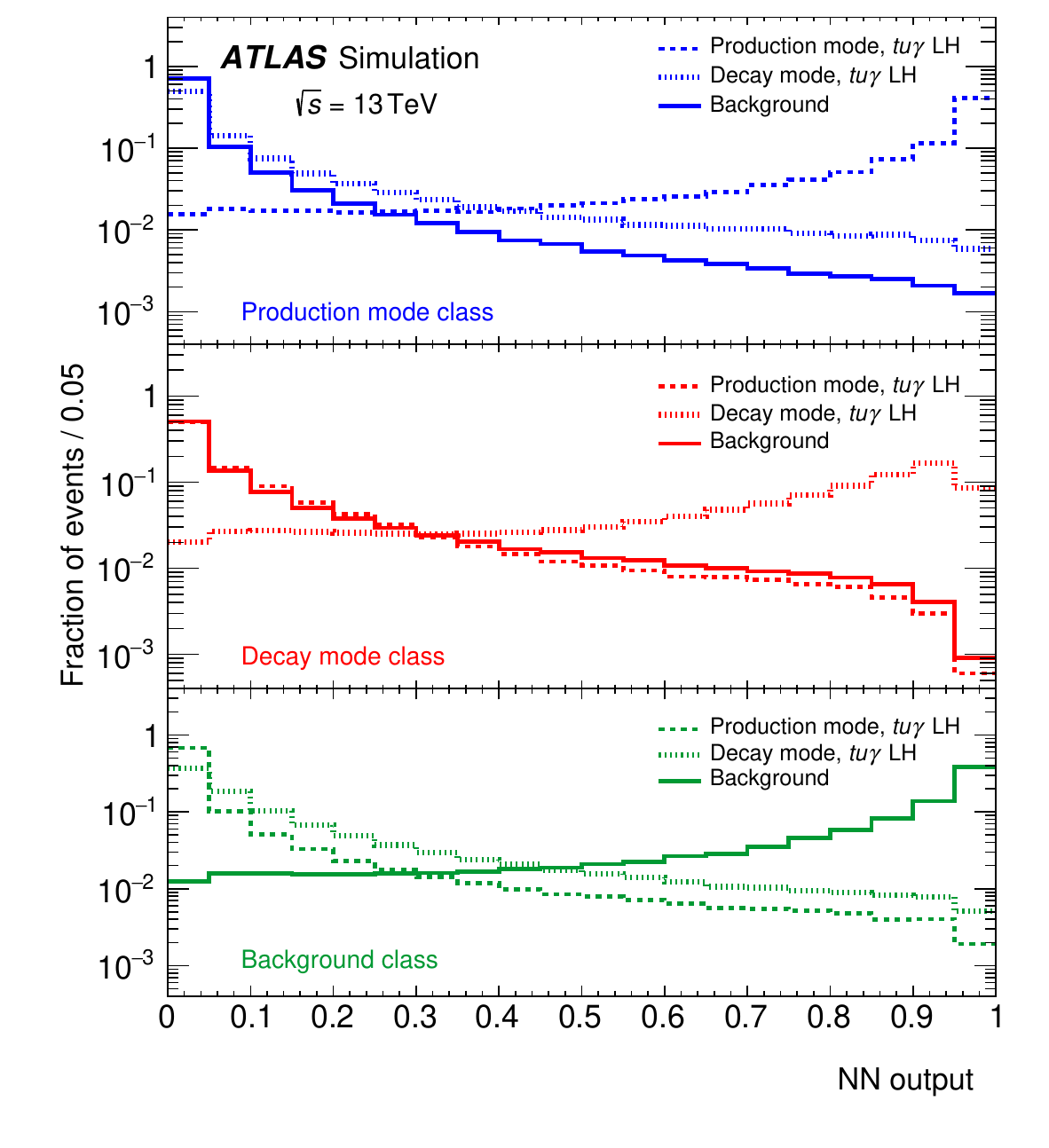}
\caption{The classifier output of the multiclass NN for the $tu\gamma$ LH coupling. The output of the FCNC production mode class is shown in blue (top), the output of the FCNC decay mode class is shown in red (middle) and the output of the SM background class is shown in green (bottom). For each output class, the distribution of the three processes is shown: FCNC production mode with the dashed line, FCNC decay mode with the dotted line and SM background with the solid line. The vertical axis represents the fraction of events predicted by the simulation.}
\label{fig:multiclass}
\end{center}
\end{figure}

The NN consists of six hidden layers with 512, 256, 128, 64, 32, 16 nodes, respectively.
The network architecture was optimised, using the expected limit without considering systematic uncertainties, to provide the best separation of the simulated FCNC signals and SM background.
The output of the NN consists of three nodes representing the three classes: the FCNC production signal, the FCNC decay signal and the SM background.
The softmax function, $f_i(x) = \exp(x_i)/\sum_{j=1}^3 \exp(x_j)$, in three dimensions for the $i$-th class is used for the activation of the output nodes.
Consequently, the target vector of the NN in the training is $(1,0,0)^\top$ for FCNC production mode events, $(0,1,0)^\top$ for the decay mode, and $(0,0,1)^\top$ for all background processes.
The NN is trained with the {\textsc Adam} optimiser~\cite{DBLP:journals/corr/KingmaB14}.
The output of the multiclass discriminator is illustrated in Figure~\ref{fig:multiclass}, showing good separation between the three output classes.
The strongest separation is achieved between the FCNC production class and the SM background class.
 
From the three-dimensional NN output, a one-dimensional discriminant, $\mathcal{D}$, is formed using
 
\begin{equation*}
\mathcal{D} = \ln{\frac{a\cdot y_{\text{prod}} + (1-a)\cdot y_{\text{dec}}}{y_{\text{bkg}}}},
\end{equation*}
 
\noindent where $y_{\text{prod(dec)}}$ represents the NN output for the FCNC production (decay) class and $y_{\text{bkg}}$ represents the NN output for the SM background class.
The discriminant $\mathcal{D}$ is inspired by the log-likelihood ratio, with an optimisable parameter $a \in (0,1)$ that changes the relative contribution of the NN outputs for the signal modes to the discriminant.
The discriminant is in the range $(-\infty, +\infty)$.
The optimal value of the parameter $a$ was found to be 0.3 for the $tu\gamma$ NN, and 0.2 for the $tc\gamma$ NN, reflecting the smaller contribution of the production mode in the case of charm-quark-initiated FCNC $t\gamma$ production.
The multiclass NN was found to outperform a simpler binary NN that discriminates only between the FCNC signal and the SM background.
The expected upper limit on the signal was found to be up to 30\% lower in the case of the multiclass neural network.
 
\FloatBarrier


\section{Systematic uncertainties}

Systematic effects may change the expected numbers of events from the signal and background processes and the shape of the fitted discriminants in the SR and the CRs.
These effects are evaluated by varying each source of systematic uncertainty by $\pm1\sigma$ and considering the resulting deviation from the nominal expectation as the uncertainty.
 
Uncertainties due to the theoretical cross-sections are evaluated by varying the cross-section by $\pm 5.6\%$ for $\ttbar$ production~\cite{Czakon:2011xx,Botje:2011sn,Martin:2009bu,Gao:2013xoa,Ball:2012cx}, by ${}^{+4.0}_{-3.4}\%\,({}^{+5.0}_{-4.5})\%$ for $t$-channel single-(anti-)top production~\cite{Kidonakis:2011wy}, by ${}^{+3.6}_{-3.1}\%\,({}^{+4.8}_{-4.3}\%)$ for $s$-channel single-(anti-)top production~\cite{Kidonakis:2010tc}, by $\pm5.3\%$ for $tW$ production~\cite{Kidonakis:2010ux}, by $\pm 5\%$ for $W$\!+jets and $Z$+jets production~\cite{ATL-PHYS-PUB-2016-003}, by $\pm30\%$ for $Z\gamma$+jets production, by $\pm50\%$ for the SM $tq\gamma$ process, and by $\pm 6\%$ for diboson production~\cite{ATL-PHYS-PUB-2016-002}.
No cross-section uncertainty is considered for $\ttbar\gamma$ and $W\gamma$+jets production, because their normalisations are determined in the fit.
 
Uncertainties due to the modelling of the signal are estimated by considering independent variations of the renormalisation and factorisation scales by factors of 2 and 0.5, but normalising the signal to the nominal cross-section.
Additionally, an uncertainty due to the choice of parton shower generator is estimated by considering the change when using an alternative MC sample that uses the \HERWIG[7.2.1]~\cite{Bahr:2008pv,Bellm:2015jjp} prediction with the H7UE set of tuned parameters~\cite{Bellm:2015jjp} and the \MMHT[lo] PDF set~\cite{Harland-Lang:2014zoa} instead of the \PYTHIA[8.212] generator.
Uncertainties due to the PDFs are estimated by using the NNPDF set of replicas.
 
For the background processes, uncertainties due to the renormalisation and factorisation scales and from the PDFs are estimated separately for each process, following the same procedure as for the signal.
An additional uncertainty due to the chosen \hdamp parameter value for the \ttbar\ process is estimated by doubling it to three times the top-quark mass.
For the $\ttbar\gamma$, \ttbar\ and single-top processes, an uncertainty is estimated from an independent variation of the A14 tune to its Var3c up and down variants~\cite{ATL-PHYS-PUB-2017-007}.
For the \ttbar\ and single-top processes, an uncertainty due to the final-state radiation modelling is estimated from an independent variation of the renormalisation scale for emissions from the parton shower by factors of 2.0 and 0.5.
In addition, a variation of the $K$-factor for decay-sample $\ttbar\gamma$ events from $1.67$ to $1.97$ is considered in order to account for uncertainties in the extrapolation to the search phase-space.
The variation is motivated by the dedicated calculation yielding the inclusive NLO $K$-factor of 1.30 used in Ref.~\cite{TOPQ-2017-14} assuming no $K$-factor for the production sample.
This uncertainty is decorrelated between the regions to account for the phase-space dependence of the inclusive $K$-factor.
The uncertainty due to the choice of parton shower generator is estimated by comparing the nominal predictions with an alternative set using \MGNLO interfaced with \HERWIG for the $\ttbar\gamma$ and SM $tq\gamma$ processes, and \POWHEGBOX[v2] interfaced with \HERWIG for the \ttbar\ and single-top processes.
The parton shower uncertainty for the $\ttbar\gamma$ and $\ttbar$ processes is fully decorrelated between the individual analysis regions to relax the constraints from these uncertainties by taking into account possible phase-space differences.
The impact on the \ttbar\ process is further split between $e\to\gamma$ and $h\to\gamma$ fakes.
Additionally, for the $\ttbar\gamma$ CR, the impact of the parton shower modelling of \ttbar\ $h\to\gamma$ fakes is split into distribution normalisation and residual uncertainties where the normalisation difference is removed.
Furthermore, an uncertainty due to the choice of generator is considered for the \ttbar\ and single-top processes by comparing the nominal prediction with a prediction from \MGNLO interfaced with \PYTHIA[8].
Finally, for the $tW$ single-top process, an uncertainty due to the overlap with the \ttbar\ process is estimated by replacing the diagram removal scheme with the diagram subtraction scheme~\cite{Frixione:2008yi}.
 
An uncertainty of 1.7\% in the integrated luminosity is considered~\cite{ATLAS-CONF-2019-021} for all processes.
The uncertainty due to pile-up is determined by varying the average number of interactions per bunch-crossing by 3\% in the simulation.
The uncertainties due to the SFs for electrons and hadrons that are misidentified as photons are determined as described in Section~\ref{sec:photonfakes}.
 
For the triggering, reconstruction, identification, and calibration of the objects, systematic uncertainties are evaluated for the following: electron and muon trigger, reconstruction, identification and isolation SFs~\cite{EGAM-2018-01,PERF-2017-01,PERF-2015-10}; photon identification~\cite{EGAM-2018-01,ATL-PHYS-PUB-2016-014} and isolation SFs; electron- and photon-energy and muon-momentum calibration and resolution~\cite{EGAM-2018-01,PERF-2015-10}; jet energy scale~\cite{PERF-2016-04} and jet energy resolution~\cite{ATL-PHYS-PUB-2015-015}; JVT SF~\cite{ATLAS-CONF-2014-018}; $b$-tagging SFs~\cite{FTAG-2018-01,ATLAS-CONF-2018-001,ATLAS-CONF-2018-006}; \met\ soft term~\cite{PERF-2016-07}.
 
The uncertainty originating from the limited number of simulated MC events is implemented via the Barlow--Beeston approach~\cite{Barlow:249779}.
Two uncertainties for each bin of the fitted distributions are considered, one for the uncertainty originating from the SM backgrounds and one for the combined production and decay FCNC signal.


\section{Results}

\begin{figure}[!ht]
\begin{center}
\includegraphics[width=0.49\columnwidth]{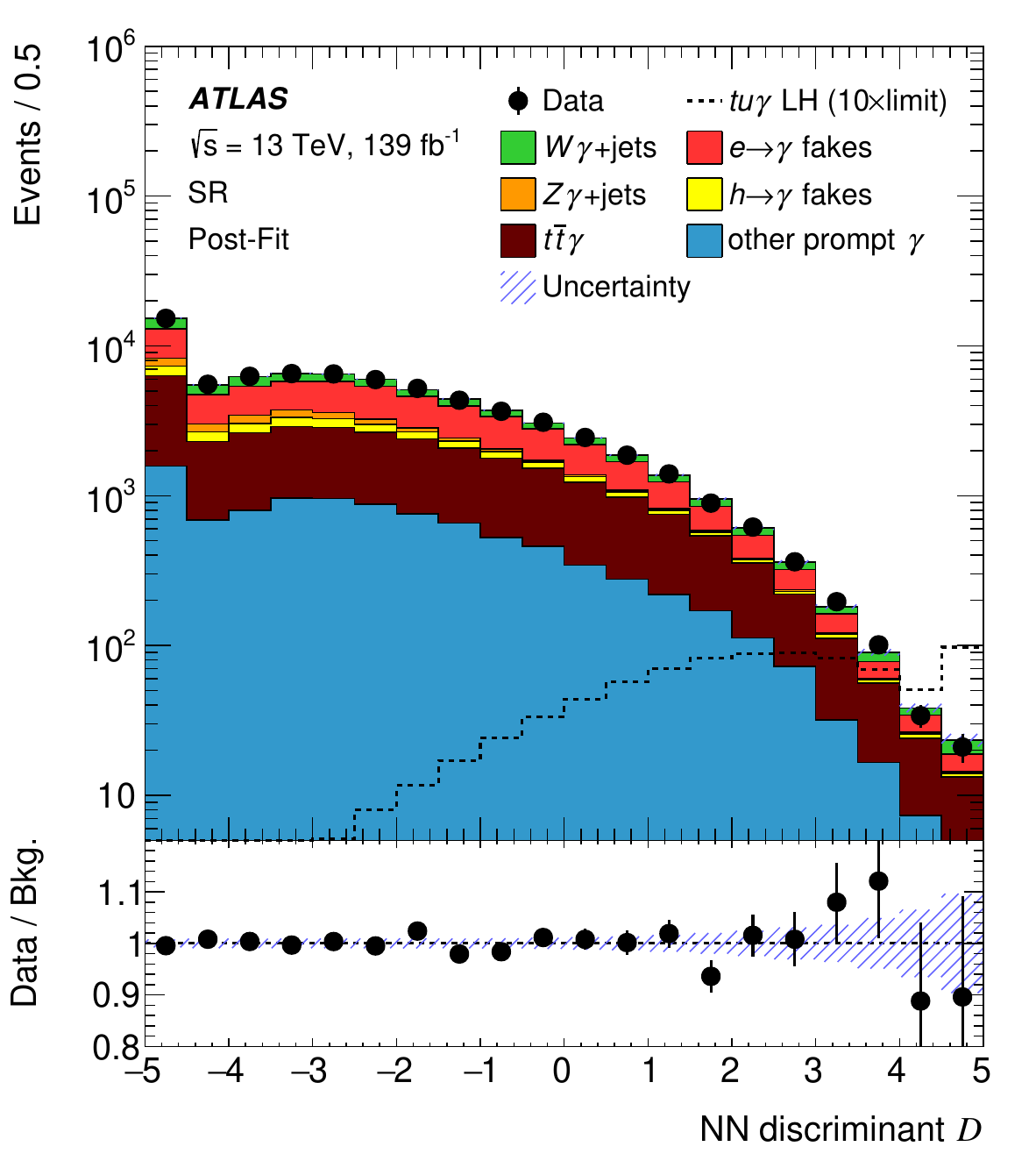} \\
\includegraphics[width=0.49\columnwidth]{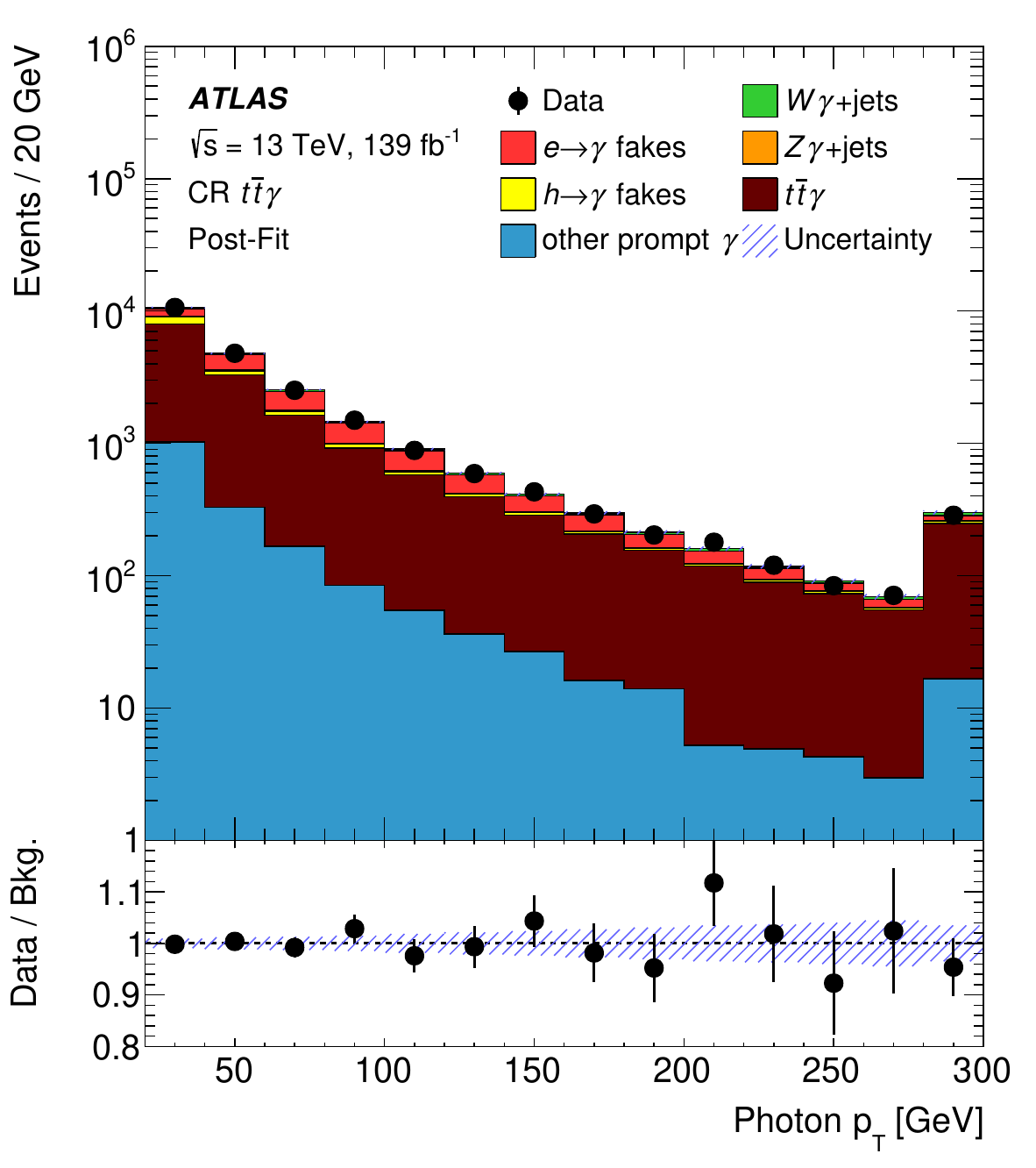}
\includegraphics[width=0.49\columnwidth]{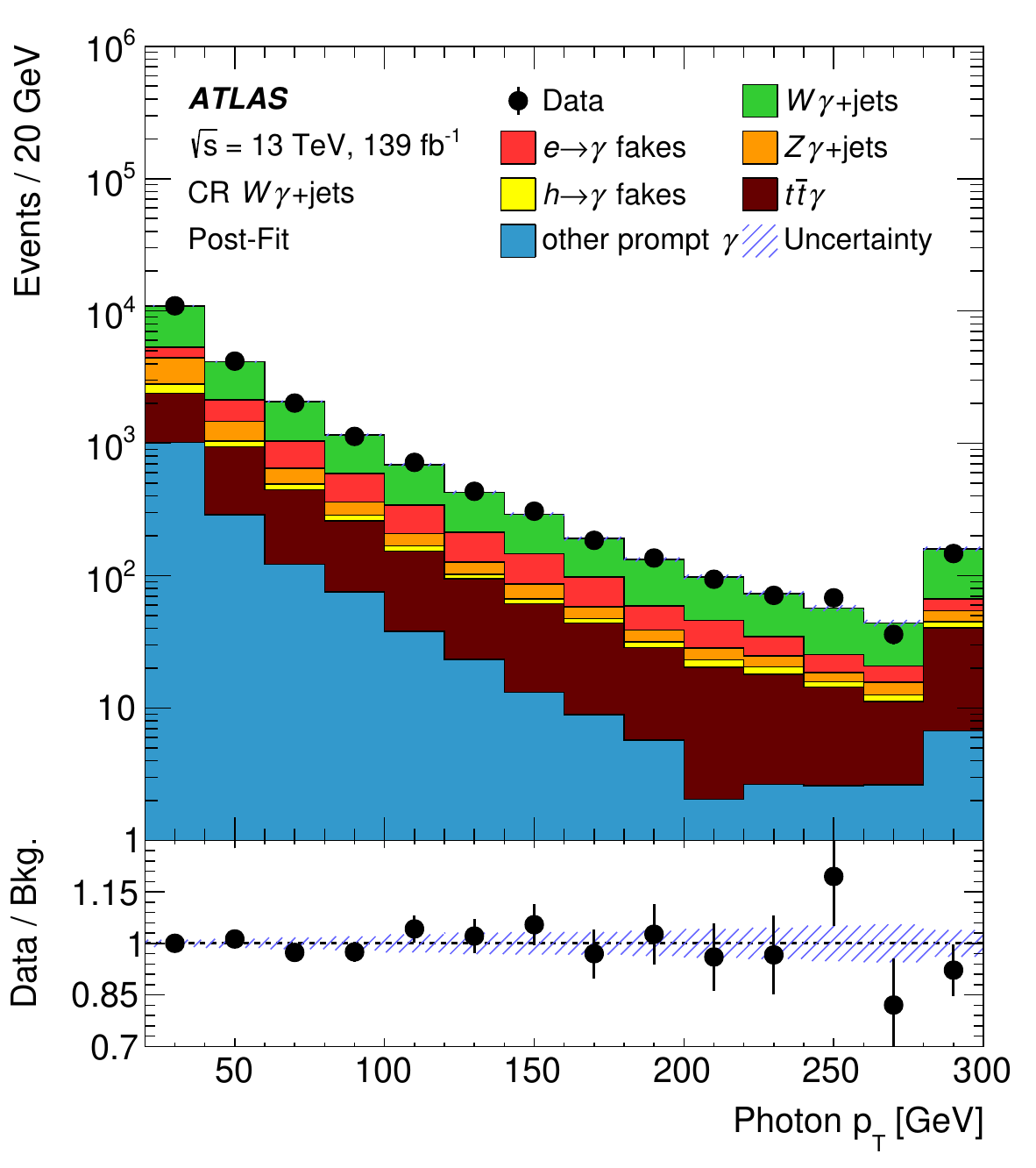}
\caption{Post-fit distributions of a background-only fit to the NN discriminant in the SR for the $tu\gamma$ coupling (top) and the photon \pt\ distribution in the $\ttbar\gamma$ (bottom left) and $W\gamma$+jet CRs (bottom right). The last bin of the distributions contains the overflow and in the case of the SR, the first bin also contains the underflow. In addition, in the SR, the expected $tu\gamma$ LH signal is overlaid for an expected number of events corresponding to the observed 95\% CL limit scaled by a factor of ten, representing BR for the $t\to u\gamma$ LH coupling of $8.5\times 10^{-5}$. The lower panels show the ratios of the data (`Data') to the background prediction (`Bkg.'). The uncertainty band includes both the statistical and systematic uncertainties in the background prediction. Correlations among uncertainties were taken into account as determined in the fit.
\label{fig:postfit}}
\end{center}
\end{figure}

The normalisations of the signal contribution and the two contributions from $\ttbar\gamma$ and $W\gamma$+jets production are obtained from a simultaneous binned profile-likelihood fit to the NN discriminant distribution in the SR and the photon \pt\ distribution in the $\ttbar\gamma$ and $W\gamma$+jets CRs, with systematic uncertainties included as nuisance parameters.
Figure~\ref{fig:postfit} shows the corresponding post-fit distribution of the NN discriminant for the $tu\gamma$ signal.
The signal is scaled to 10 times the observed upper limit on the BR for the $t\to u\gamma$ LH coupling, corresponding to a BR of $8.5 \times 10^{-5}$.
The qualitative features of these distributions are similar for the $tc\gamma$ NN.
 
\begin{figure}[ht!]
\begin{center}
\includegraphics[width=0.49\columnwidth]{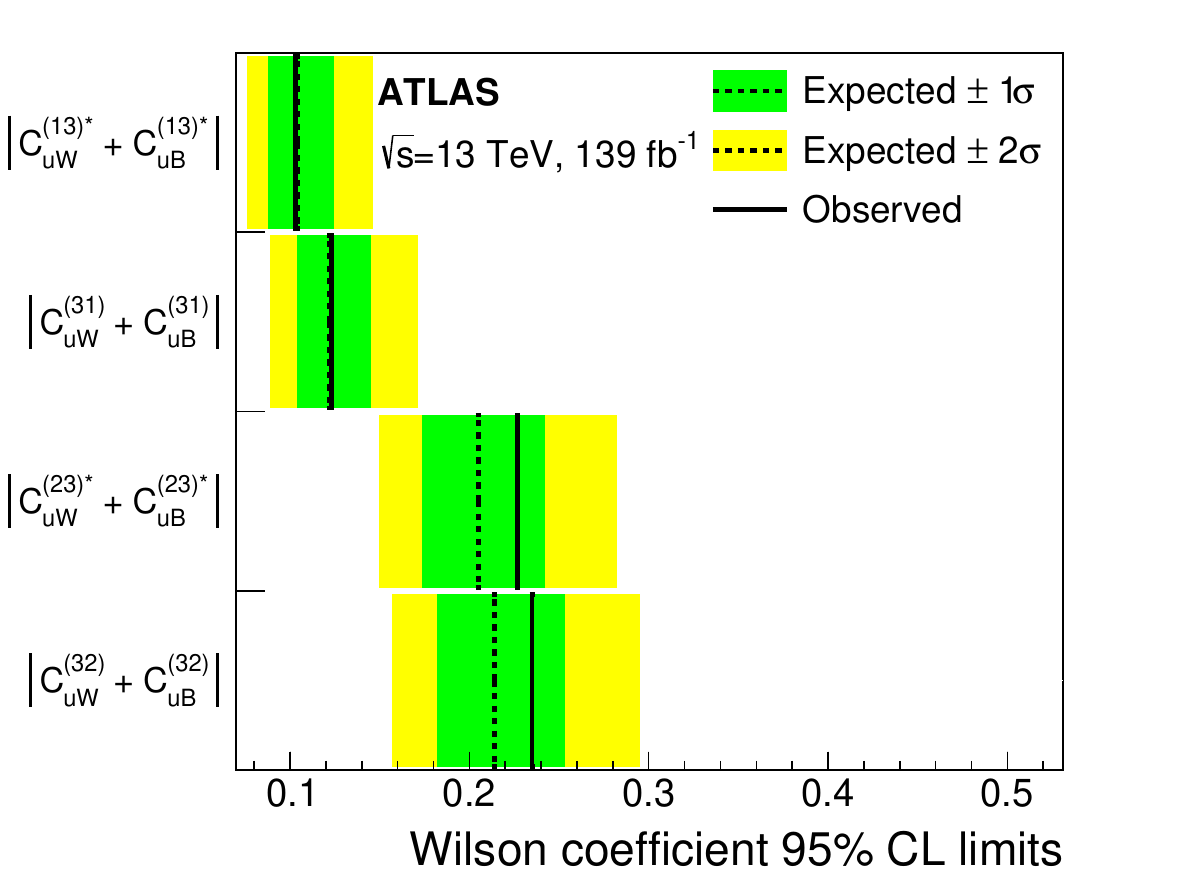}
\includegraphics[width=0.49\columnwidth]{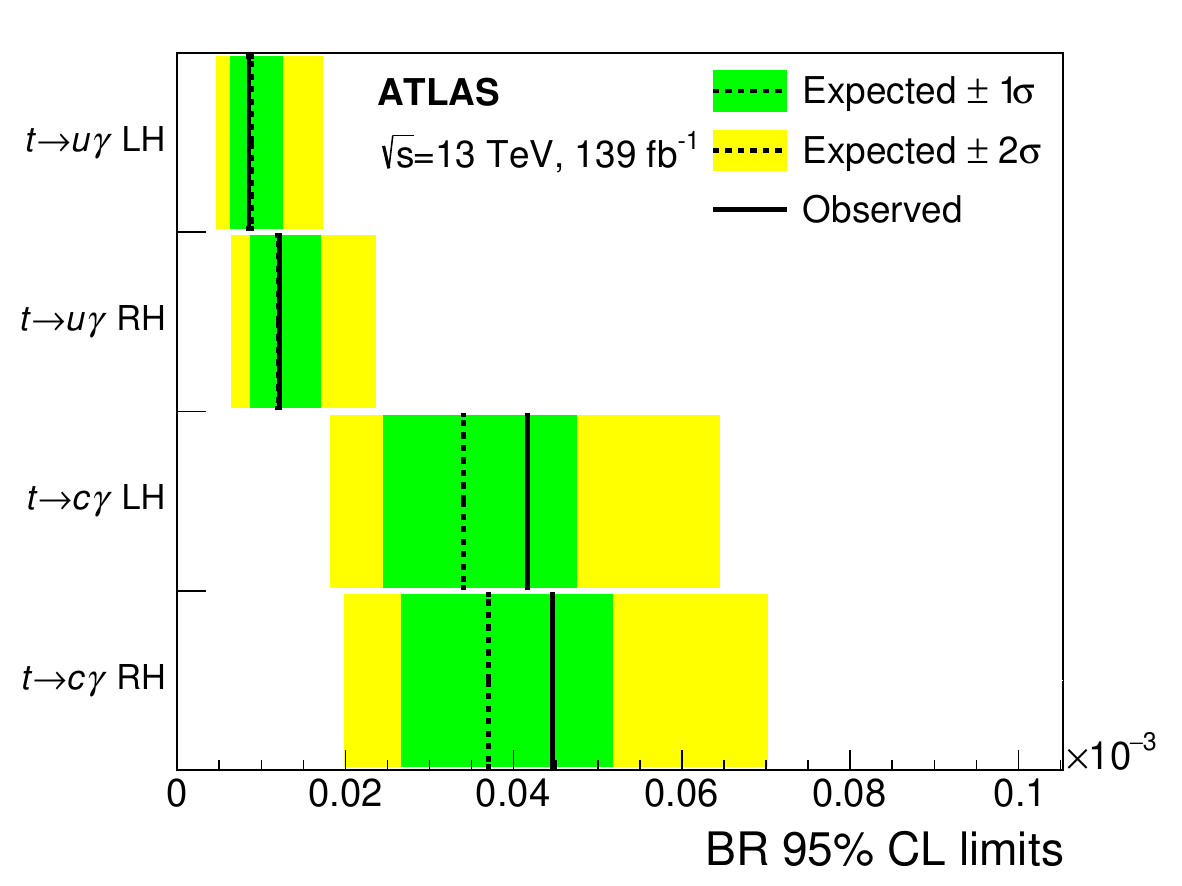}
\caption{The 95\% CL upper limits on the effective coupling constants (left) and BRs (right). The expected (observed) limits are shown with the dotted (solid) lines. The green (yellow) bands represent one (two) standard deviations for the limits. The scale of new physics is set to $\Lambda = 1~\tev$.
\label{fig:limits}}
\end{center}
\end{figure}

The data and SM predictions agree within uncertainties and no significant FCNC contributions are observed.
From the 95\% confidence level (CL) upper limits on the signal contribution, derived using the $\mathrm{CL_{s}}$ method~\cite{Read:2002hq}, the corresponding limits on the effective coupling parameters are calculated~\cite{Cowan:2010js}.
From these, limits on the BRs are also derived.
The background contributions from $\ttbar\gamma$ and $W\gamma$+jets production are scaled by normalisation factors estimated to be $1.00 \pm 0.10$ and $1.15 \pm 0.15$, respectively, from the fit for the $tu\gamma$ LH coupling.
The normalisation values determined in the fit for the $tc\gamma$ LH coupling are $0.97 \pm 0.10$ for the $\ttbar\gamma$ contribution and $1.16 \pm 0.15$ for $W\gamma$+jets.
The normalisation values are similar for the fits with the RH FCNC couplings.
The K-factors for the $\ttbar\gamma$ decay-sample are fitted to the values of about 1.57 for the SR, about 1.61 for the $\ttbar\gamma$ CR and about 1.73 for the $W\gamma$+jets CR.
The smallest post-fit uncertainty on the K-factor is seen in the $\ttbar\gamma$ CR, where the uncertainty is found to be about 0.24.
The observed and expected 95\% CL limits on the effective coupling strengths, the BRs and the production cross-sections are presented in Table~\ref{tab:Results/Limits} as well as in Figure~\ref{fig:limits}.
The fitted normalisations of the FCNC $tc\gamma$ signal are larger than zero, although consistent with zero within one standard deviation.
Thus, the observed 95\% CL limits for the $tc\gamma$ couplings are larger than the expected 95\% CL limits.
The dominant source of uncertainty in the signal contribution for the $tu\gamma$ couplings is the statistical uncertainty.
All systematic uncertainties worsen the limit by only about 20\%.
For the $tc\gamma$ couplings, the effect of the systematic uncertainties is larger, worsening the limit by about 40\%, but the statistical uncertainties also play an important role.
The different importance of the systematic uncertainties on the observed limits between $tu\gamma$ and $tc\gamma$ couplings result in a difference in the improvement in the limits compared to the previous results~\cite{TOPQ-2018-22} for the two couplings.
The sources of systematic uncertainty with the largest impact on the estimated signal contribution depend on the coupling studied.
For the $tu\gamma$ couplings, the dominant systematic uncertainties arise from the cross-section uncertainty of the SM $tq\gamma$ process, the uncertainty in the electron--photon energy scale and the decay $K$-factor uncertainty in the $\ttbar\gamma$ process.
For the $tc\gamma$ couplings, the dominant systematic uncertainties originate from the cross-section uncertainty of the SM $tq\gamma$ process, the uncertainty in the hadron-to-photon SFs and the limited number of simulated events for the SM backgrounds.
No nuisance parameters are significantly pulled or constrained.
The difference in the improvement for the $tu\gamma$ and $tc\gamma$ coupling with respect to the previously published result~\cite{TOPQ-2018-22} stems mainly from the different relative contribution of the production and decay modes between the couplings.
The efficiency for the decay mode with higher jet multiplicity is particularly increased compared to the previous result, increasing the improvement for the $tc\gamma$ coupling limits.
 
\begin{table}[!hb]
\centering
\caption{The 95\% CL limits on the effective coupling constants, BRs and production cross-sections. The expected limits with their uncertainties, representing one standard deviation, as well as observed limits are shown. The scale of new physics is set to $\Lambda = 1~\tev$.}
\label{tab:Results/Limits}
{\renewcommand{\arraystretch}{1.3}
\begin{tabular}{cccc}
\toprule
Observable & Coupling & Expected & Observed \\
\midrule
$|C_{\mathrm{uW}}^{(13)*} + C_{\mathrm{uB}}^{(13)*}|$ & $t\rightarrow u\gamma$ LH & $0.104^{+0.020}_{-0.016}$ & 0.103 \\
$|C_{\mathrm{uW}}^{(31)\phantom{*}} + C_{\mathrm{uB}}^{(31)\phantom{*}}|$ & $t\rightarrow u\gamma$ RH & $0.122^{+0.023}_{-0.018}$ & 0.123 \\
$|C_{\mathrm{uW}}^{(23)*} + C_{\mathrm{uB}}^{(23)*}|$ & $t\rightarrow c\gamma$ LH & $0.205^{+0.037}_{-0.031}$ & 0.227 \\
$|C_{\mathrm{uW}}^{(32)\phantom{*}} + C_{\mathrm{uB}}^{(32)\phantom{*}}|$ & $t\rightarrow c\gamma$ RH & $0.214^{+0.039}_{-0.032}$ & 0.235 \\
\midrule
BR ($t\to q\gamma$) $\times 10^{5}$ & $t\rightarrow u\gamma$ LH & $0.88^{+0.37}_{-0.25}$ & 0.85 \\
BR ($t\to q\gamma$) $\times 10^{5}$ & $t\rightarrow u\gamma$ RH & $1.20^{+0.50}_{-0.33}$ & 1.22 \\
BR ($t\to q\gamma$) $\times 10^{5}$ & $t\rightarrow c\gamma$ LH & $3.40^{+1.35}_{-0.95}$ & 4.16 \\
BR ($t\to q\gamma$) $\times 10^{5}$ & $t\rightarrow c\gamma$ RH & $3.70^{+1.47}_{-1.03}$ & 4.46 \\
\midrule
$\sigma(pp \to t\gamma)$ [fb] & $t\rightarrow u\gamma$ LH & $11.2^{+4.7}_{-3.1}$ & 10.9 \\
$\sigma(pp \to t\gamma)$ [fb] & $t\rightarrow u\gamma$ RH & $15.3^{+6.4}_{-4.3}$ & 15.6 \\
$\sigma(pp \to t\gamma)$ [fb] & $t\rightarrow c\gamma$ LH & $6.2^{+2.5}_{-1.7}$ & 7.6 \\
$\sigma(pp \to t\gamma)$ [fb] & $t\rightarrow c\gamma$ RH & $6.8^{+2.7}_{-1.9}$ & 8.2 \\
\bottomrule
\end{tabular}
}
\end{table}
 
\FloatBarrier


\FloatBarrier
 
\section{Conclusion}

A search for flavour-changing neutral currents (FCNCs) in events with one top quark and a photon is presented using 139~fb$^{-1}$ of $\sqrt{s} = 13$~\TeV\ $pp$ collision data collected with the ATLAS detector at the LHC.
Events with a photon, an electron or muon, exactly one $b$-tagged jet, missing transverse momentum and possibly additional jets are selected.
The contribution from events with electrons or hadrons that are misidentified as photons is estimated from simulation with data-driven corrections, and the two main background processes with a prompt photon are estimated with control regions.
Multiclass neural networks are used to distinguish events with FCNCs in the production process, events with FCNCs in the decay process, and background events.
The outputs are combined into a one-dimensional discriminant to search for the FCNC signal.
The data are consistent with the background-only hypothesis.
Limits are set on the strength of effective operators that introduce a left- or right-handed flavour-changing $tq\gamma$ coupling with an up-type quark $q$.
The 95\% CL bounds on the corresponding branching fractions for the FCNC top-quark decays are $\mathcal{B}(t\rightarrow u\gamma) < 0.85 \times 10^{-5}$ and $\mathcal{B}(t\to c\gamma) < 4.2 \times 10^{-5}$ for a left-handed $tq\gamma$ coupling, and $\mathcal{B}(t\to u\gamma) < 1.2 \times 10^{-5}$ and $\mathcal{B}(t\to c\gamma) < 4.5 \times 10^{-5}$ for a right-handed coupling.
The obtained limits are the most stringent to date and improve on the previous ATLAS limits by factors of 3.3 to 5.4.
The improvements in the limits originate mainly from considering events with more than one jet, the optimisation of the signal separation and the increased integrated luminosity.


\section*{Acknowledgements}


We thank CERN for the very successful operation of the LHC, as well as the
support staff from our institutions without whom ATLAS could not be
operated efficiently.
 
We acknowledge the support of
ANPCyT, Argentina;
YerPhI, Armenia;
ARC, Australia;
BMWFW and FWF, Austria;
ANAS, Azerbaijan;
CNPq and FAPESP, Brazil;
NSERC, NRC and CFI, Canada;
CERN;
ANID, Chile;
CAS, MOST and NSFC, China;
Minciencias, Colombia;
MEYS CR, Czech Republic;
DNRF and DNSRC, Denmark;
IN2P3-CNRS and CEA-DRF/IRFU, France;
SRNSFG, Georgia;
BMBF, HGF and MPG, Germany;
GSRI, Greece;
RGC and Hong Kong SAR, China;
ISF and Benoziyo Center, Israel;
INFN, Italy;
MEXT and JSPS, Japan;
CNRST, Morocco;
NWO, Netherlands;
RCN, Norway;
MEiN, Poland;
FCT, Portugal;
MNE/IFA, Romania;
MESTD, Serbia;
MSSR, Slovakia;
ARRS and MIZ\v{S}, Slovenia;
DSI/NRF, South Africa;
MICINN, Spain;
SRC and Wallenberg Foundation, Sweden;
SERI, SNSF and Cantons of Bern and Geneva, Switzerland;
MOST, Taiwan;
TENMAK, T\"urkiye;
STFC, United Kingdom;
DOE and NSF, United States of America.
In addition, individual groups and members have received support from
BCKDF, CANARIE, Compute Canada and CRC, Canada;
PRIMUS 21/SCI/017 and UNCE SCI/013, Czech Republic;
COST, ERC, ERDF, Horizon 2020, ICSC-NextGenerationEU and Marie Sk{\l}odowska-Curie Actions, European Union;
Investissements d'Avenir Labex, Investissements d'Avenir Idex and ANR, France;
DFG and AvH Foundation, Germany;
Herakleitos, Thales and Aristeia programmes co-financed by EU-ESF and the Greek NSRF, Greece;
BSF-NSF and MINERVA, Israel;
Norwegian Financial Mechanism 2014-2021, Norway;
NCN and NAWA, Poland;
La Caixa Banking Foundation, CERCA Programme Generalitat de Catalunya and PROMETEO and GenT Programmes Generalitat Valenciana, Spain;
G\"{o}ran Gustafssons Stiftelse, Sweden;
The Royal Society and Leverhulme Trust, United Kingdom.
 
The crucial computing support from all WLCG partners is acknowledged gratefully, in particular from CERN, the ATLAS Tier-1 facilities at TRIUMF (Canada), NDGF (Denmark, Norway, Sweden), CC-IN2P3 (France), KIT/GridKA (Germany), INFN-CNAF (Italy), NL-T1 (Netherlands), PIC (Spain), ASGC (Taiwan), RAL (UK) and BNL (USA), the Tier-2 facilities worldwide and large non-WLCG resource providers. Major contributors of computing resources are listed in Ref.~\cite{ATL-SOFT-PUB-2023-001}.


\clearpage
 
\printbibliography
 
\clearpage
 
\begin{flushleft}
\hypersetup{urlcolor=black}
{\Large The ATLAS Collaboration}

\bigskip

\AtlasOrcid[0000-0002-6665-4934]{G.~Aad}$^\textrm{\scriptsize 101}$,
\AtlasOrcid[0000-0002-5888-2734]{B.~Abbott}$^\textrm{\scriptsize 119}$,
\AtlasOrcid[0000-0002-7248-3203]{D.C.~Abbott}$^\textrm{\scriptsize 102}$,
\AtlasOrcid[0000-0002-2788-3822]{A.~Abed~Abud}$^\textrm{\scriptsize 36}$,
\AtlasOrcid[0000-0002-1002-1652]{K.~Abeling}$^\textrm{\scriptsize 55}$,
\AtlasOrcid[0000-0002-8496-9294]{S.H.~Abidi}$^\textrm{\scriptsize 29}$,
\AtlasOrcid[0000-0002-9987-2292]{A.~Aboulhorma}$^\textrm{\scriptsize 35e}$,
\AtlasOrcid[0000-0001-5329-6640]{H.~Abramowicz}$^\textrm{\scriptsize 150}$,
\AtlasOrcid[0000-0002-1599-2896]{H.~Abreu}$^\textrm{\scriptsize 149}$,
\AtlasOrcid[0000-0003-0403-3697]{Y.~Abulaiti}$^\textrm{\scriptsize 116}$
% ...
\bigskip
\\

% ...
$^{29}$Physics Department, Brookhaven National Laboratory, Upton NY; United States of America.\\
% ...
$^{35}$$^{(a)}$Facult\'e des Sciences Ain Chock, R\'eseau Universitaire de Physique des Hautes Energies - Universit\'e Hassan II, Casablanca;$^{(b)}$Facult\'{e} des Sciences, Universit\'{e} Ibn-Tofail, K\'{e}nitra;$^{(c)}$Facult\'e des Sciences Semlalia, Universit\'e Cadi Ayyad, LPHEA-Marrakech;$^{(d)}$LPMR, Facult\'e des Sciences, Universit\'e Mohamed Premier, Oujda;$^{(e)}$Facult\'e des sciences, Universit\'e Mohammed V, Rabat;$^{(f)}$Institute of Applied Physics, Mohammed VI Polytechnic University, Ben Guerir; Morocco.\\
$^{36}$CERN, Geneva; Switzerland.\\
% ...
$^{55}$II. Physikalisches Institut, Georg-August-Universit\"{a}t G\"ottingen, G\"ottingen; Germany.\\
% ...
$^{101}$CPPM, Aix-Marseille Universit\'e, CNRS/IN2P3, Marseille; France.\\
$^{102}$Department of Physics, University of Massachusetts, Amherst MA; United States of America.\\
% ...
$^{116}$Department of Physics, New York University, New York NY; United States of America.\\
% ...
$^{119}$Homer L. Dodge Department of Physics and Astronomy, University of Oklahoma, Norman OK; United States of America.\\
% ...
$^{149}$Department of Physics, Technion, Israel Institute of Technology, Haifa; Israel.\\
$^{150}$Raymond and Beverly Sackler School of Physics and Astronomy, Tel Aviv University, Tel Aviv; Israel.\\
% ...
$^{a}$ Also Affiliated with an institute covered by a cooperation agreement with CERN.\\
$^{b}$ Also at Borough of Manhattan Community College, City University of New York, New York NY; United States of America.\\
$^{c}$ Also at Bruno Kessler Foundation, Trento; Italy.\\
$^{d}$ Also at Center for High Energy Physics, Peking University; China.\\
$^{e}$ Also at Centro Studi e Ricerche Enrico Fermi; Italy.\\
$^{f}$ Also at CERN, Geneva; Switzerland.\\
% ...

\end{flushleft}


\end{document}